\documentstyle[amssymb,preprint,aps]{revtex}

\begin{document}
\title{{\bf The Predictive Power of }$R_{0}${\bf \ in an Epidemic
Probabilistic System}}
\author{D. Alves$^{1,2}$, V. J. Haas$^{3}$ and A. Caliri$^{2\thanks{%
To whom correspondence should be sent}}$}
\address{1) Laborat\'{o}rio Interdisciplinar de Computa\c{c}\~{a}o Cient%
\'{\i}fica,\\
Faculdades COC\\
Rua Abrah\~{a}o Issa Hallack, 980\ -- 14096-160 \ \ Ribeir\~{a}o Preto, SP\\
\--\\
Brazil\\
2) Departamento de F\'{\i}sica e Qu\'{\i}mica, FCFRP - Universidade de\\
S\~{a}o\\
Paulo\\
Av. do Caf\'{e} S/N \ -- 14040-903 Ribeir\~{a}o Preto, SP \ \ -- \ Brazil.\\
3) Departamento de Patologia, FM - Universidade de S\~{a}o Paulo\\
Av. Dr. Arnaldo, 455 \ -- 01246-903 S\~{a}o Paulo, SP \ -- Brazil.}
\date{May, 2002}
\maketitle

\begin{abstract}
An important issue in theoretical epidemiology is the epidemic threshold
phenomenon, which specify the conditions for an epidemic to grow or die out.
In standard (mean-field-like) compartmental models the concept of the {\it %
basic reproductive number}, $R_{0}$, has been systematically employed as a
predictor for epidemic spread and as an analytical tool to study the
threshold conditions. Despite the importance of this quantity, there are no
general formulation of $R_{0}$ when one considers the spread of a disease in
a generic finite population, involving, for instance, arbitrary topology of
inter-individual interactions and heterogeneous mixing of susceptible and
immune individuals. The goal of this work is to study this concept in a
generalized stochastic system described in terms of global and local
variables. In particular, the dependence of $R_{0}$ on the space of
parameters that define the model is investigated; it is found that near of
the ``classical'' epidemic threshold transition the uncertainty about the
strength of the epidemic process still is significantly large. The
forecasting attributes of $R_{0}$ for a discrete finite system is discussed
and generalized; in particular, it is shown that, for a discrete finite
system, the pretentious predictive power of $R_{0}$ \ is significantly
reduced.
\end{abstract}


\newpage \qquad

\section{Introduction}

It is nowadays recognized that the phenomenon of health-disease in human
communities only may be understood by considering complex and dynamic
inter-relations among several factors operating simultaneously in multiple
spatiotemporal and organizational scales. In fact, the healthy and sick
individual suffers uninterruptedly the effects of the microbiological
evolution, the antropogenic environmental and ecosystem stress and many
others misdeeds resulting from socioeconomic inequalities. Therefore, it is
not surprising to find out the proliferation of a myriad of methodological
tools employed during the development of the epidemiological research.

Among this methodological mosaic the mathematical and computer (or
simulation) modeling of communicable and infectious disease comes as a
hypothetical-deductive approach whose scope consists primarily in
understanding and manipulating, {\it a priori} and to predictive purposes,
the underlying mechanisms behind the origin and diffusion of epidemic
events. As a matter of fact, the attempt of understanding in what conditions
pathogenic agents (once invaded a host population) could establish
themselves as an infection (the transmission of pathogens from one host to
another) resulted in the development of one of the most important and
thoroughly discussed concepts in infectious disease modeling as early as in
the beginning of the last century, namely the {\it epidemic threshold}.
Thus, in writings of R. Ross (1909) \cite{Ross} the so-called {\it mosquito
theorem}{\em \ }was the first recognition of a quantitative threshold
deducing that it was not necessary to eliminate mosquitoes totally in order
to eradicate malaria. Two decades later would testify the publication of the
classic Kermack-McKendrick's (1927) \cite{Kermack} paper that definitely
consolidated the threshold concept in epidemiologic literature. In this
deterministic $SIR$ model ($S$ stands for susceptibles, $I$ for infected,
and $R$ for removed) an epidemic process is considered to evolve only when
the density of susceptible individuals is greater than a threshold value $%
S_{c}$. Bartlett (1957) \cite{Bartlett}, based in a large amount of
collected data of disease incidence in industrialized countries introduced
thirty years later another expression linking microbial invasion and
threshold pa\-ra\-me\-ters: the {\it critical community size}, that could
explain the fade-out patterns of measles epidemics.

However, the inherent individual heterogeneity and probabilistic local
nature of interindividual relationships has been traditionally neglected in
state-variable models like this; in fact, in this population level approach
all behavioral and individual variability are diluted into the
intercompartmental rates and densities or number of mean individuals ---as $%
S $, $I$ or $R$ compartments--- described in terms of partial or ordinary
differential equations. Nevertheless, it was subsequently possible to
express the epidemic threshold in a way perhaps much more intuitive when the
focus changed to consider the infected host or the parasite itself, instead
of looking at the density or number of susceptible. In this perspective the
threshold condition that determines whether an infectious di\-se\-a\-se will
spread in a susceptible population has been described through the so-called 
{\em basic reproduction\ number}{\it \ }or also denominated as {\it basic
reproductive rate}, commonly denoted by ${\cal R}_{0}$ \cite{Keeling2000}.
For microparasites such as viruses or bacteria it may be biologically
understood as the average number of secondary cases produced or caused by
one infected individual during its entire infectious period in a completely
susceptible po\-pu\-la\-tion. Thus, the intrinsically individual based
perspective of this threshold concept should not be underestimated since the
reproduction number can link the inside-host evolutionary or pathogenic
dynamic (microscale) and transmission process at population level
(macroscale). From a purely deterministic point of view it appears
intuitively evident that if ${\cal R}_{0}\geq 1$ the pathogen can
undoubtedly establish itself in a host population and, at least, an endemic
regime will settle down. But this is a short-sighted prediction since,
specially to directly transmitted disease in finite populations, the
mechanisms that ensure the maintenance of the parasite within a community
depends critically on the way as the individuals interact one another,
sometimes unforeseeable.

In this work we analyze limitations of the predictive power of the{\em \ }$%
R_{0}$ parameter (as classically formulated) for the spread of a disease:
Alternatively to population level approach and state-variable models, \
stochastic inter-individual interactions are also used and its implications
on the predictive attributes of the {\it basic} {\it reproduction}{\em \ }%
{\it number} ${\cal R}_{0}$ are studied through a simplified model: a
lattice based model including infectious period in that individual
interactions are straightforwardly described in terms of global $(\,\Gamma
\,)$ and local $(\Lambda )$ variables, which in turn can be tuned out to
simulate respectively the populational mobility and geographical
neighborhood contacts.

The remainder of this paper is organized as follows. In the next section it
is presented a general formalism to the evolution of a population invaded by
an infection.\ The formalism is then applied in section {\em 3, }where
concepts involving \ ${\cal R}_{0}$ and the threshold phenomenon are
discussed in order to define an invasion criterion for the infection and
evaluation of \ ${\cal R}_{0}$.\ The results are discussed in section {\em 4}%
. Although this work will be mainly concerned on ${\cal R}_{0}$ as a
function of the model's parameters, the formalism presented in what follows
can be applied to study a variety of \ epidemic scenarios.

\section{The Model System}

Consider a discrete dynamical system (discrete space and discrete time)
where a population of $N$ individuals is distributed on the sites of a
toroidal lattice ${\bf M}=\{m_{ij}\}$ ---with $i$ and $j$ varying from $1$
to $L$ ($N=L\times L$). Each individual site $m_{ij}$ is assigned to receive
three personal specific attributes: {\em (1)} a spatial address or lattice
position $(i,j)$; ${\em (2)}$ a set of three possible {\it status, }namely, $%
s,i$ and $r,$ specifying a clinic disease stage of each particular
individual, which represent, respectively, the conditions of {\it susceptible%
} (subject to be infected by a contagious agent), {\it infectious}
(effectively transmitter of contagious agents) and {\it removed} (recovered
or immune); and finely {\em (3)} an infectious period $\tau $, specifying
how many units of time an infected individual can propagate the contagious
agent. Note that $\sum s+\sum i+\sum r=N$,\ with $N$ constant.

\ Such a system is suitable mainly for describing a single epidemic in a
closed system (no birth or migration). The choice of such reduced model,
however, is not far-fetched because, as already mentioned above, the main
interest here involves only very short period of time, so that the dynamics
of host births, migration, etc., are largely irrelevant. \cite{Anderson}.
The dynamic evolution of the population is described, step-by-step, by a set
of {\it a priori} stated interaction rules, and assumes that each new
configurational state of the system (described here by the geographical
address $(i,j)$ of each individual and by the instantaneous number of \
susceptibles $S(t)$, infectives $I(t),$ and removed individuals $R(t)$ )
depends only on its previous state. Hence, for the present purpose the
spread of the disease in the population is considered as being governed by
the following rules:

\begin{enumerate}
\item Any susceptible individual may become infected with a probability $%
p_{S}$. An infected susceptible becomes infective after an average latency
time $\tau _{l}$ (assumed here as $\tau _{l}=0,$ without lost of generality).

\item Infectives are removed deterministically from the system (becoming
immune) after an infectious period $\tau $, that for simplicity is
considered as constant for all infected individuals.

\item Once in the removed class the individual participate only passively in
the spreading of the infection (eventual topological blocking) by a period
of immunity greater than the complete epidemic process.
\end{enumerate}

During one time step, the three preceding rules are applied synchronously to
all sites in the lattice; the present model, therefore, can be viewed as a
simple two-dimensional cellular automaton.\ Actually, it is an adaptation of
automata network to standard $SIR$ models for studying the spread of
infectious diseases.

In this work, the probability $p_{S}$, which is intended to be probability
per unit of time, is taken as the superposition of the local and global
influences, in order to unify the individual-based (contacts among nearest
neighbors) and the standard mean-field (homogeneously interacting
population) approaches. Therefore, one assumes that disease transmission
occurs with a total infection probability $p_{S}$ written as 
\begin{equation}
p_{S}=\Gamma p_{G}+\Lambda p_{L},  \label{eq0}
\end{equation}
where the pre-factors $\Gamma $ and $\Lambda $ are weight parameters tuning
the short (cluster formation) and long-range (mean-field type) interactions;
it is also required that $\Gamma +\Lambda =1$ in order to satisfy the
probabilistic requirement $0\leqq p_{S}\leqq 1$.

The global influence $p_{G}$ amounts to the probability of a susceptible to
become infective due to the ubiquity of $I(t)$ infected individuals
(mean-field). So one can expect that in the limit of large $N$ ($%
N\rightarrow \infty )$, in each time step, any susceptible may become
infected with probability 
\begin{equation}
p_{G}=\frac{\rho }{N}\sum\limits_{\{k,l\}}\delta _{i,\sigma (k,l)}
\label{eq1}
\end{equation}
where $0\leq \rho \leq 1$ is one of the model parameters: it limits the
maximum value of $p_{G}$ and is related to the intrinsic mobility of the
population; the sum sweeps all lattice sites $\{k,l\},$ and $\delta
_{i,\sigma (k,l)}$ is the Kronecker delta function which assumes the value
``one'' when the state{\it \ }$\sigma $ of the site $(k,l)$ corresponds to
the infectious state $i$, and ``zero'' otherwise ($\sigma (k,l)$ can be $s,i$
or $r)$. Actually, the sum in the Equation 2 just counts the instantaneous
number of infectious individuals $I(t)$ in the population.

On the other hand, the local term $p_{L}=p_{L}(i,j)$ is the probability of a
susceptible individual (located at the site $(i,j))$ contracting infection
due to $n$ infectives first and second neighbors ( $0\leq n\leq 8$ is a
integer number corresponding to all possible combinations of $(i+\xi
,\,j+\xi )$, with $\xi =0,1,-1)$. \ Therefore, let $\lambda \in \left[ 0,1%
\right] $ be the probability of a particular susceptible when just one of
its neighbors is infective. Hence, $(1-\lambda )^{n}$ will be the
probability for not contracting the disease when exposed to $n$ infectives.
Therefore, the chance of he (or she) contracting the disease in a unit of
time is\cite{Cardy}

\begin{equation}
p_{L}=1-\left( 1-\lambda \right) ^{n}.  \label{eq2}
\end{equation}
Thus, when $\lambda =1$ the infection spreads deterministically, with $8$
nearest neighbors to any infective being infected (the choice for equipotent
first and second neighbors was adopted because the use of only the four
nearest neighbors is unduly restricting).

The expression for $p_{G}$ is a convenient and simple way for describing the
populational mobility. It is based on the {\it mass action law}, borrowed
from the chemistry, and gets new meaning here under the perspective of
pairwise spatially disordered interactions through the population elements.
In this sense, it is a result of the small-word effect, and so became a
particular version of the small-word lattice of Watts and Strogatz \cite%
{Strogatz}

This simple approach allows to study in great detail the dynamical behavior
of the model in the full space of control parameters $\lambda $ and $\rho $,
and the local and global balance pre-factors $\Lambda $ and $\Gamma $.
Therefore, the system is governed\ by $p_{S}$ \ (Eq.1) and $\tau $, and its\
temporal evolution is determined by updating the lattice synchronously at
each time step through the application of the three rules above.

\section{${\cal R}_{0}$ and The Threshold Phenomenon}

The probability $p_{L}$ as in the Eq.(3) \cite{Grassberger}, \cite{Cardy},
and in a number of alternative forms\cite{Claudia}, has been employed in the
analogy between percolation and epidemic. Since that the critical value $%
p_{c},$ in which random clusters grow to infinite size, is know (analytic or
numerically) for any lattices, $p_{c}$ may be used as a powerful general
criterion for ``epidemic spread''\cite{Grassberger2}, \cite{Claudia}.
However, due to the traditional importance of the concept of $R_{0}$ in the
epidemic scenario, this threshold is generalized for finite discrete
systems, as described above, in order to show its relevance for an
intrinsically individual based perspective of the problem.

The overall structure of the model presented here shows the interplay of two
types of transmission mechanisms by assuming that each infectious individual
interacts strongly (physically) with their few susceptible neighbors, and
uniformly and weakly with each particular susceptible in the population of
susceptibles. Thus, the local mode of transmission $p_{L}$ incorporates the
individual-based component from the perspective of the susceptible
individuals, the actual (physical) contacts that each susceptible
experiences, and the global probability $p_{G},$ due to intrinsic
populational mobility, which may be viewed as resulting of a mean-field
(discrete) approach, in the sense that the disease transmission to each
susceptible individual also depends on the instantaneous total number of
infectious individuals in the population.

To better appreciate\ the consequences of this formulation, it is firstly
run si\-mu\-la\-ti\-ons for the extreme values of the tuning pre-factors
through the procedure described above. These cases allow to recover the two
modes of transmission in its pure form corresponding to the ({\it i})
homogeneous mixing approximation (mean field), when $\Gamma =1$, and to the (%
{\it ii}) percolation process (the transmission occurs by localized
individual contact), when $\Lambda =1$.\ Furthermore, it is considered the 
{\it damage} $\Delta I$ on a susceptible population due to just one infected
individual ($I(0)=1$) landing in a totally susceptible population ($S(0)=N-1$
and $R(0)=0$) during the infectious period $\tau ;$ to calculate $\Delta I$
it is considered only the number of new infected individuals in the
population after $\tau $ time steps, ignoring infections from the victims of
these first infected individual (operationally, it is enough to consider the
latency $\tau _{l}>\tau $, that is, a latent period of infection greater
then the infectious period). The Figures 1a and 1b show, respectively, the
si\-mu\-la\-ti\-on results for $\Delta I$ as function of \ $\rho $ for $%
\Gamma =1$, and the behavior of $\Delta I$ \ as a function of the contact
probability $\lambda $ for $\Lambda =1$ (that is, $\Gamma =0)$; the system
size considered in most of the simulations presented here was $L=100$
(population size $N=L\times L=10^{4}),$ although some extra different sizes $%
L/2$ and $2L$) were also used in order to verify finite size effects. $%
\Gamma =1$ and $\Lambda =1$ are the two limiting cases usually taken as
references in studying the effect on the system when both mechanisms are
superposed; the amount $\Delta I$ \ is obtained after $\tau =10$ time steps
(covering exactly the infectious period) and was estimated as an average
over $31$ independent simulations (what is equivalent to verify the
establishment of infection on $31$ distinct populations with the same
pattern of contacts among the individuals).

The linear pattern observed for $\Delta I$ vs $\rho $ \ means that the
present stochastic approach reproduces qualitatively the classical basic
reproductive number ${\cal R}_{0}$ if one identifies $\Delta I$ as the
average number of secondary cases that an infectious individual causes.
Indeed, the linear \ relation $\Delta I=[\frac{\rho }{N}\,S(0)]\,\,\tau $ \
fits pretty well the data shown in Figure 1a, and so one may consider that
infectives make contacts at a mean rate $[\frac{\rho }{N}\,S(0)]\,$
throughout an infectious period of length $\tau $ (note that for large
enough populations $\frac{\rho }{N}\,S(0)\rightarrow \rho $). On the other
hand, when $\Lambda =1,$ the amount $\Delta I$ represents ${\cal R}_{0}$ for
the case where individuals interact only with their spatial nearest
neighbors, and so its values saturates at $\Delta I=8$ for $\lambda \gtrsim
0.3$. \ For each particular run, significant fluctuations on $\Delta I$ are
observed (mainly for smaller $N)$ but averaging over 31 runs is enough to
smooth considerably the curves, as shown in Figure 1.

Before to proceed through the application of the present formulation, some
comments regarding the definition of the ${\cal R}_{0}$ are in order. The
basic reproduction number has been widely used as a predictor parameter
conceived to indicate the epidemic potential of a pathogen once it has
introduced in a totally susceptible population. In fact, to deterministic
and continuous (in space and in time) population-system models the future
fate of an infectious agent has been expressed through the threshold
condition. Accordingly, when ${\cal R}_{0}>1$, infections can invade a
totally susceptible population and persist; if ${\cal R}_{0}<1$, the disease
then dies out and can not establish itself. To the special condition ${\cal R%
}_{0}=1$, there is an endemic regime in that the unique initial infectious
case reproduces subsequently just one infectious secondary case and son on.

This assumption in modelling of the establishment of an infection (which is
possibly wrong) \cite{Mollison} will be partially preserved here to have the
classical treatment as a reference but, indeed, to capture more realistic or
probable practical situations is of interesse that the ``first analytical
look'' at a population be considered when the epidemic process is already in
course. For instance, at the initial time $t_{0}$ one may consider the
arbitrary situation in that $I(t=t_{0})>>1$ at the same time that the number
of removed individuals is also large, and then ask the question: What is the
value for the {\it reproduction}{\em \ }{\it number} in this case? To answer
this question one may generalize the concept of ${\cal R}_{0}$ as the
normalized average number ${\cal R}(t_{0};\tau )$ of secondary cases
(reproductive ratio) about the time $t_{0},$ due to $I(t_{0})$ infectious
present in the population at $t=t_{0},$ through the following expression 
\begin{equation}
{\cal R}(t_{0};\tau )=\frac{\sum_{t_{0}}^{t_{0}+\tau }\left\langle
\sum_{\{k,l\}_{s}}(\Gamma p_{G}+\Lambda p_{L})\right\rangle _{n}}{I(t_{0})},
\label{Rinicial}
\end{equation}
where the brackets means an average on a set of $\ n$ independent runs in
the time interval $[t_{0},\,t_{0}+\tau ],$ and the sum over $\{k,l\}$ sweeps
all sites occupied by individuals in the {\it status} $s$ (susceptibles).
Note that all the instantaneous extensive and intensive conditions of the
population, at any arbitrary time $t_{0},$ are all taken into account, as
for example, the sites in the removed {\it status} randomly scattered
through the population (acting as epidemic shield protectors), and the set
remaining infectious time $\tau (k,l;$ $t_{0})$ for each individual in the 
{\it status} $i$ located at the site $(k,l).$ These conditions certainly
affect the epidemic process and the progression of the epidemic process
depends in some how on the {\it reproduction}{\em \ }{\it number'}s value,%
{\it \ }(that is, if ${\cal R}(t_{0};\tau )>1$ or $<1).$ \ But, as already
mentioned above, the initial condition $I(t=t_{0})=1$ will be deliberately
used in the present work in order to maintain the original intention of
comparing the traditional deterministic definition of the basic reproductive
ratio ${\cal R}_{0}$ with the present stochastic approach.

In order to infer how the intrinsic stochastic nature of the epidemic
process affects the predictive attributes of ${\cal R}_{0},$ the concept of 
{\it epidemic pro\-ba\-bi\-li\-ty} $P_{E}$ is introduced. Numerically it is
estimated directly from the simulation experiments based on the algorithm of
the previous section. Indeed, it is just given by the ratio $P_{E}=n_{e}/n,$
where $n_{e}$ is the number of runs in that at least one susceptibles was
infected during the infectious period, and $n$ is the total number of runs
or experimental populations. The probability $P_{E}$ may be expressed as
function of the mean reproduction number ${\cal R}_{0}$, which also is
determined from the same simulation experiments by using the Equation 4
above. In the Figure 2 it is shown the resulting $P_{E}$ as a function of $%
{\cal R}_{0}$ with $\Gamma =\Lambda =0.5$ and $\rho $ and $\lambda $ varying
in the interval $(0\,-\,0.2]$. The large number of scattered points in the
graph, mainly at larger ${\cal R}_{0}$, is an intrinsic aspect of this graph
due to the fact that in the parameter space $(\rho ,\lambda )$ there are
different combinations of $\rho $ and $\lambda $ resulting in approximately
the same values for ${\cal R}_{0}$, as it is illustrated in Figure 3.
Therefore only the stochastic component of such scattering of points may be
reduced by increasing the number of runs used in the averaging procedure.

The amount $P_{E}$ tends to saturates at $P_{E}\simeq 1$ when the value of $%
{\cal R}_{0}$ is sufficient large (${\cal R}_{0}\gtrsim 3$), so that the
epidemic spread in the population almost always is observed. Furthermore,
the results showed in the Figure 2 means that only for large enough ${\cal R}%
_{0}$ (actually ${\cal R}_{0}>3$) one can be sure about an epidemic
development in the population, while that, even for ${\cal R}_{0}<1$ there
is still a possibility to have an epidemic spread. Therefore, from the
epidemic control perspective, reducing the effective reproductive number to
a level below one, upon vaccination, for instance, could be a potential
problem of strategy since that for ${\cal R}_{0}\lesssim 1$ in about $60\%$
of events this strategy will fail, that is, an epidemic process should be
established with chance of $60\%$ for ${\cal R}_{0}\cong 1,$ under the
conditions of the present model. More pointedly, despite the claim of the
threshold criterium, it is improbable to recognize (using only standard
census data) the imminence of any epidemic disaster if the system is near to
the threshold region.\cite{Haas} The more accurate (although frustrating)
criterium is to realize that, irrespective the value of ${\cal R}_{0}$ that
the level of vaccination forces, there is always a chance (even thought
small) of the disease re-invading the population.

The same system size $N=10^{4}$ was employed in order to get all the results
discussed above. However, in order to verify eventual effect of the system
size on the results, two extra systems were considered, namely a smaller $%
N=4^{-1}\times 10^{4}$ and a bigger $N=4\times 10^{4}$ system, but no
significant difference was found. Clearly fluctuations are smaller for
larger systems mainly because the chance of nucleation of closer cluster due
to the global term $\Gamma p_{G}$ decreases with the system size $N,$
reducing then the chance of the magnification effect of the local term $%
\Lambda p_{L}$ on eventual clusters located nearly enough each other. The
Figure 4a shows for $P_{E}$ vs ${\cal R}_{0}$ (in the interval $0<{\cal R}%
_{0}\leq 2)$ for two different system sizes; note that the size effect is
pronounced only on the second moment (dispersion of the data) of the
distribution of $P_{E}$ for each ${\cal R}_{0}$. More precisely, the Figure
4b shows the normalized standard deviation (relative error) $\sigma _{R_{0}}$
as function of ${\cal R}_{0}$ for the larger $4\times 10^{4}$ system. A
decreasing $1/{\cal R}_{0}-$ like behavior for the relative error is a
consequence of the averaging of integer quantities$,$ that is: ${\cal R}%
_{0}=(0\times n_{0}+1\times n_{1}+2\times n_{2}+3\times n_{3}+$ $\cdot \cdot
\cdot )/\eta ;$ where $\eta =n_{0}+n_{1}+n_{3}+$ $\cdot \cdot \cdot ,$ and $%
n_{k}$ is the number of experiments in which exactly $k$ susceptibles were
infected.

Finally, the numerical equivalence between ${\cal R}_{0}^{\prime }$
estimated by an analytical approximation and $\ {\cal R}_{0}$ calculated by
simulation is verified. For this purpose ${\cal R}_{0}^{\prime }$ is
considered in the limit of large populations ($N\rightarrow \infty )$ by
taking the mean number of susceptible infected by just one infective during
its infectious period $\tau $, through the following direct expression 
\begin{equation}
{\cal R}_{0}^{\prime }=\left\{ \Gamma \left[ \frac{\rho }{N}S(0)\right]
+\Lambda \lbrack \lambda \,8]\right\} \,\tau .
\end{equation}
The Figure 5 shows the parametric graph of ${\cal R}_{0}^{\prime }$ vs $%
{\cal R}_{0}$ where they are calculated, respectively, by Equation (5) above
and by simulation using the proposed probabilistic approach represented in
Equation (4), with $I(t_{0}$ $=0)=1$. \ Strong correlation between the two
ways for estimating the {\it basic} {\it reproduction}{\em \ }{\it number}
is kept only for values of $\rho $ and $\lambda $ not too large (${\cal R}%
_{0}\lesssim 2)$ because during the time $\tau ,$ the local term that
composes ${\cal R}_{0}$ (Eq.4) may change from zero up to eight, $\ $while
this limit is not present in the Equation (5). However, that is enough in
order to validate the conclusions about the predictive attributes of $%
P_{E}=P_{E}({\cal R}_{0})$ because\ ${\cal R}_{0}^{\prime }$ and ${\cal R}%
_{0}$ are numerically equivalent: the result given by Equation 5, although
intuitive, follows from a stochastic representation of the classical SIR
model \cite{Aiello}.

\bigskip

\section{Final comments}

In this paper a stochastic version of the original $SIR$ model (involving
only single epidemics) was introduced with the main purpose of to
characterize and re-interpret the conditions for the establishment of an
epidemic in a population through the concept of {\it basic reproduction
number }(${\cal R}_{0}$). A peculiar characteristic of the present approach
is the assumption that the probability of a susceptible individual become
infective is a superposition of the local and global influences. Using as
initial configuration just one single infected individual in a fully
susceptible population, condition frequently used to define ${\cal R}_{0}$,
it was demonstrated that the discrete character of a finite population
reduces the pretentious predictability of the threshold criteria, and so it
is, indeed, an incomplete predictive tool since that, irrespective to the
value of ${\cal R}_{0}$, an epidemic has a finite probability to establish
itself, due the inherent stochastic nature of any finite epidemic system.

Indeed, more consistent derivation of $R_{0}$ has been tried, even though
using the same classical deterministic approach, due to the too widely
estimate obtained to $R_{0},$ which in recent applications for the smallpox
have varied from $R_{0}=1.5$ to 
\mbox{$>$}
20 \cite{Gani}. Rather than just a caricature of the original formulation of 
${\cal R}_{0}$, the approach presented in this paper may be viewed as a
simpler and generic alternative for investigating the spread of diseases in
a population, which may greatly facilitate the analysis of a number of
distinct epidemic scenarios. Particularly, a system with increasing
to\-po\-lo\-gi\-cal complexity can be easily tackled. For example, one may
consider the practical situation in that, at an arbitrary initial time $%
t_{0},$ the population has already many infectious individuals (that is, $%
I(t_{0})>>1)$, and also many immunes scattered through the population
(working as epidemic shield) and then try to answer the question: What is
the value for the {\it reproduction}{\em \ }{\it number} in this case? \ 

Finally, as a major challenge that this ``microscopic'' approach can \
handle, one may think on the possibility of incorporating in the traditional
definition of ${\cal R}_{0}$ the underlying evolutionary dynamics of the
pathogenic agent. This view is in contrast with the standard epidemiological
models, which tend to use a constant absolute parasite fitness ${\cal R}_{0}$%
. However, more detailed considerations on the investigation of this avenue
of research is left for a future contribution.

\bigskip

\bigskip

ACKNOWLEDGMENTS: The authors thank E. Massad for fruitful discussions and
supporting. Part of this work was supported by Funda\c{c}\~{a}o de Amparo 
\`{a} Pesquisa do Estado de S\~{a}o \ Paulo (FAPESP): \ Proc. 00/00570-1 and
00/03465-4

\bigskip

\begin{center}
\newpage FIGURE\ CAPTIONS
\end{center}

FIGURE 1: Average damage $\Delta I$ due to just one infectious individual on
the susceptible population $S=N-1,$ for two extreme cases . [{\it a}]-- $%
\,\,\,\Gamma =1$: the amount $\Delta I$ changes linearly with the intrinsic
mobility $\rho $, as can be expect from Equation (2). \ [{\it b}%
]--\thinspace\ $\,\,\Lambda =1$: the amount $\Delta I$ increases rapidly
with the infection probability $\lambda $ due to local (physical) contact,
and saturates at $\Delta I=8$ for $\lambda \gtrsim 0.3$, as one can infer
from Equation (3).

\bigskip

FIGURE 2: The epidemic probability as a function of average reproduction
number ${\cal R}_{0}$. The tuning pre-factor are fixed at $\Gamma =\Lambda
=0.5$, and the parameter $\rho $ and $\lambda $ are choosing from the
interval $\left[ 0,0.2\right] $. For ${\cal R}_{0}\simeq 1$ epidemics are
observed in about $60\%$ of the events (in a population of size $N=10^{4}$).

\bigskip

FIGURE 3 {\it Reproduction}{\em \ }{\it number} ${\cal R}_{0}$ as function
of the model parameters ($\rho $, $\lambda )$ obtained by averaging over 100
independent realizations. Each strip, identified by a different gray tone,
corresponds to a range of value for ${\cal R}_{0}$ according to: white, $%
0\leq {\cal R}_{0}<1;$ light gray, $1\leq {\cal R}_{0}<2;$ and so on. At the
limit of very large populations ($N\rightarrow \infty )$ the slope $\alpha $
(dotted lines), which roughly delimitates each region, can be obtained using
Equation (5) --see text; giving $\alpha =\frac{\Gamma }{8\Lambda }$.
Therefore, in [{\it a}] $\Gamma =\Lambda =0.5,$ so $\alpha =-0.125;$ and in [%
{\it b}] $\Gamma =0.9$ and $\Lambda =0.1,$ giving $\alpha =-1.125$, whose
values are closely reproduced by the results.

\bigskip

FIGURE 4: [{\it a}] The epidemic probability $P_{E}$ vs ${\cal R}_{0}$ for
two systems: $N=4^{-1}\times 10^{4}$ (open circles) and $N=4\times 10^{4}$
(dark circles) smaller fluctuations for the larger system is the most
significative difference. [{\it b}]- The relative error decreases as ${\cal R%
}_{0}$ increases; for ${\cal R}_{0}\simeq 1$ \ the absolute error is of the
same magnitude of ${\cal R}_{0}$ as a consequence of the averaging on \
``zeros'' and ``ones'', mainly.

\bigskip

FIGURE 5: Numerical equivalence between ${\cal R}_{0}^{\prime }$ estimated
by an analytical approximation and $\ {\cal R}_{0}$ calculated by simulation.

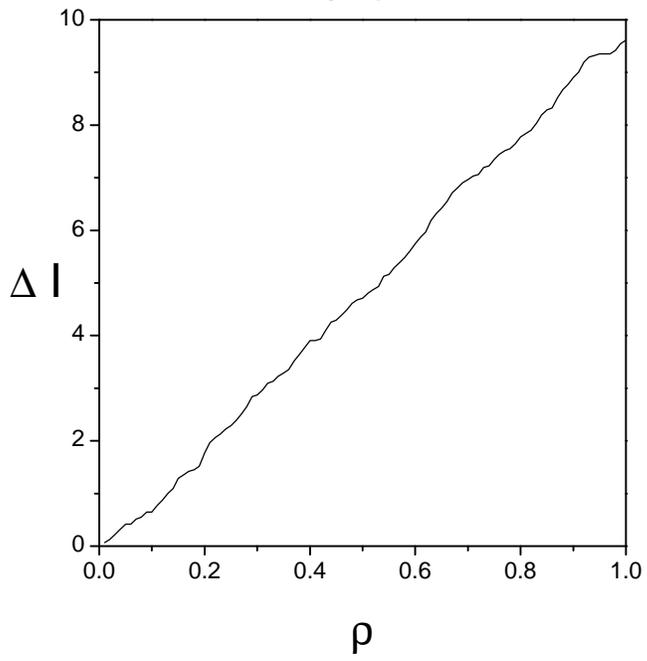
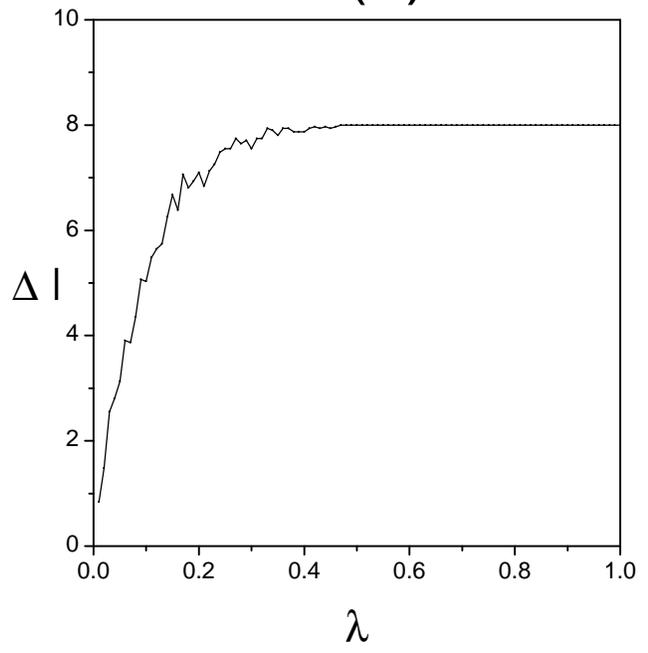
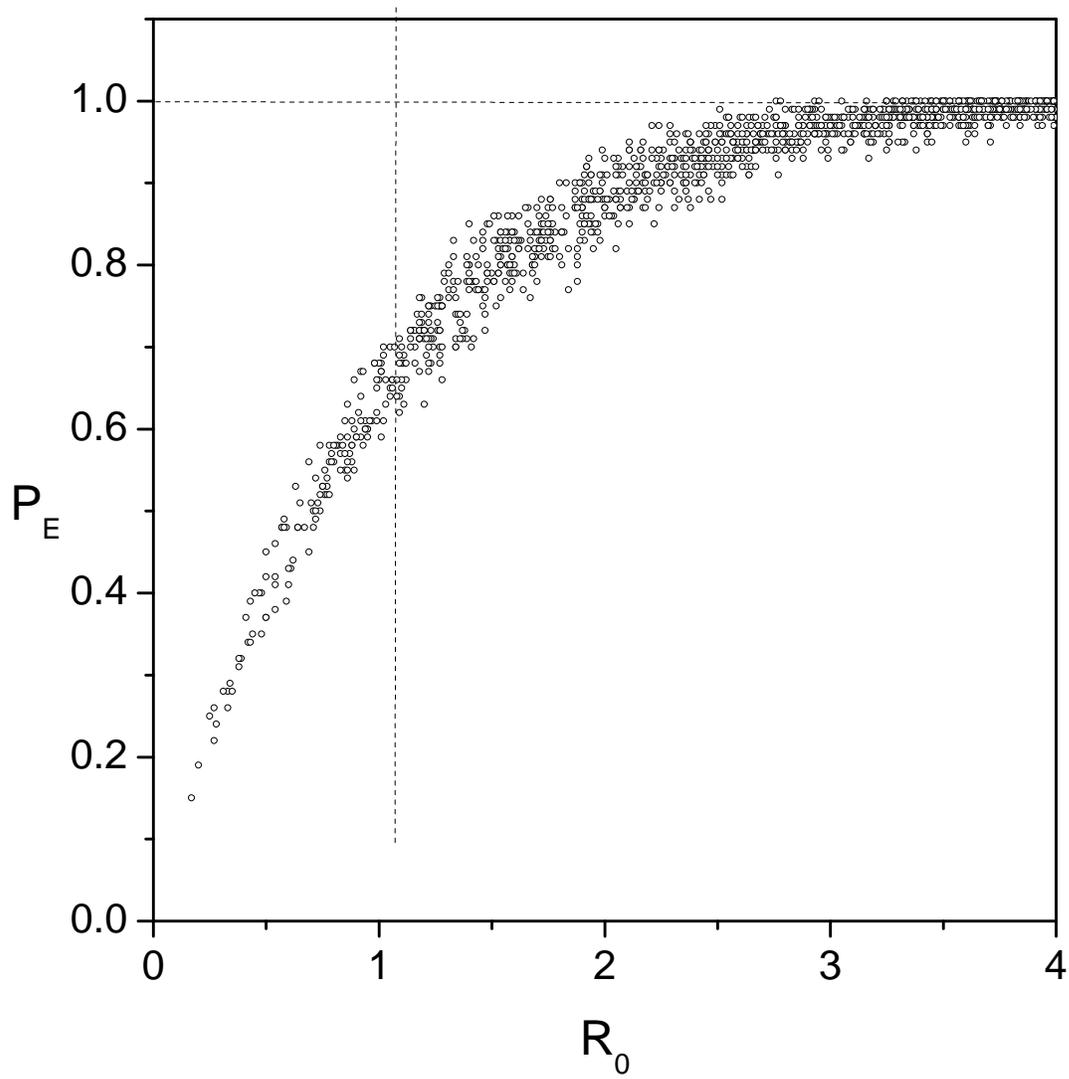
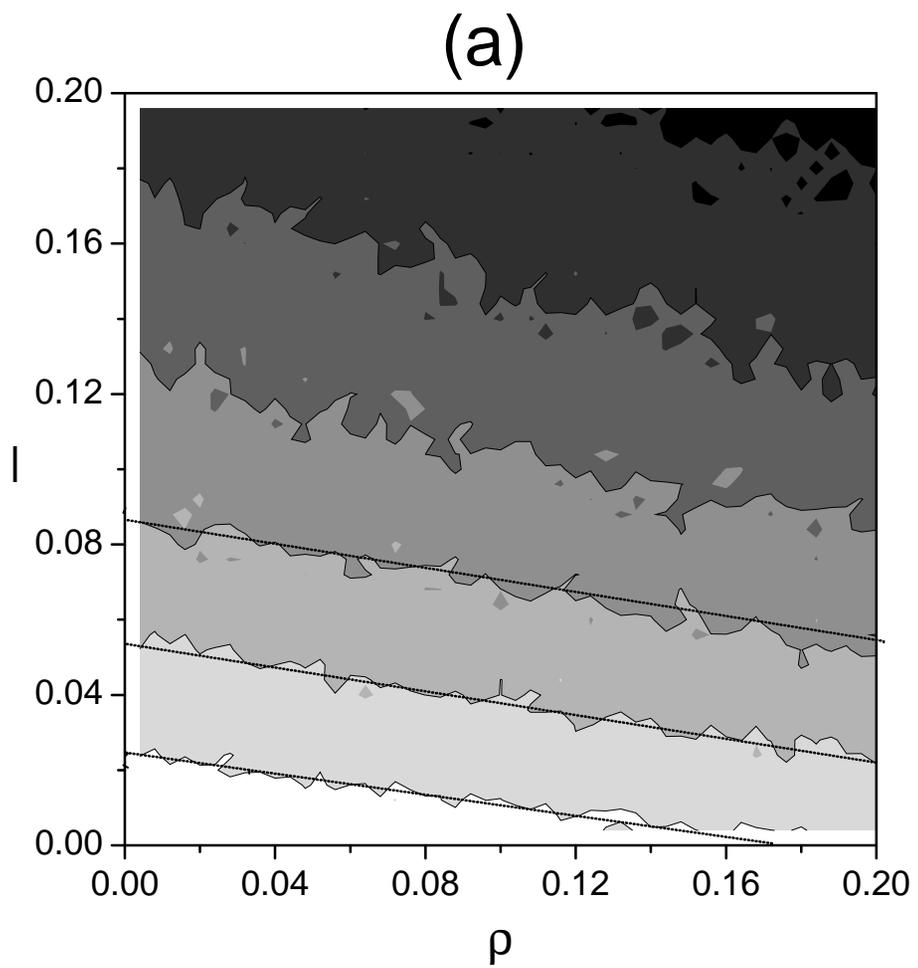
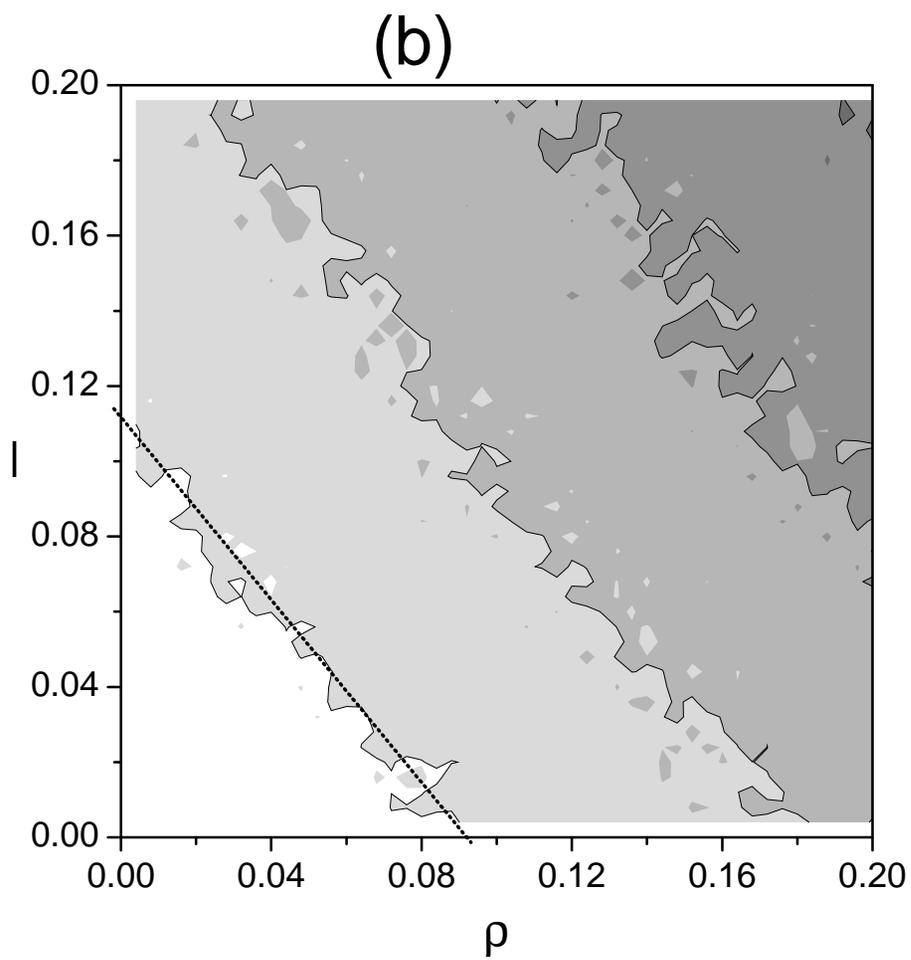
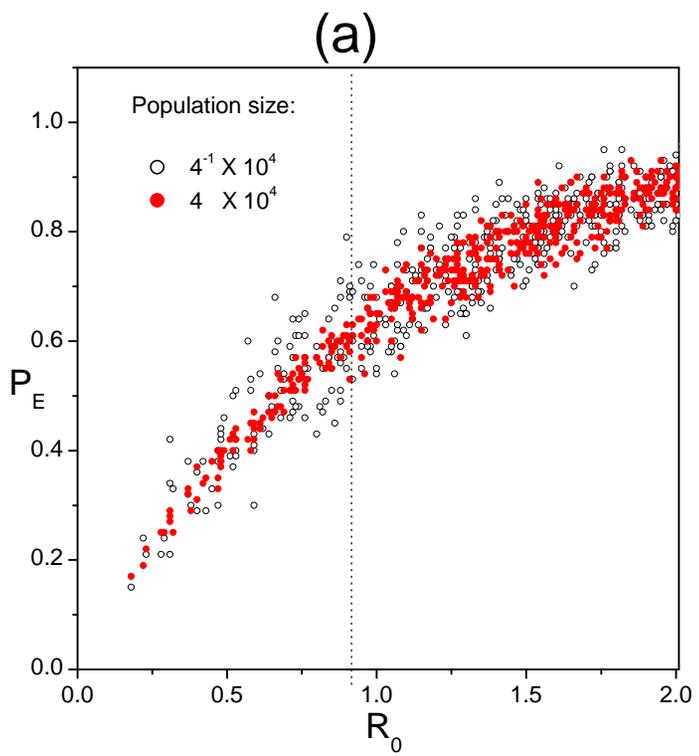
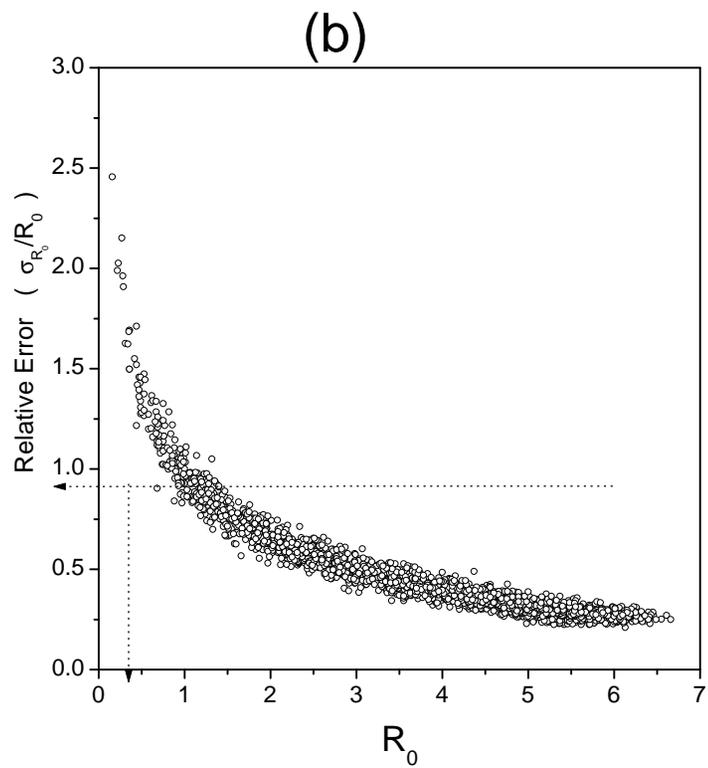
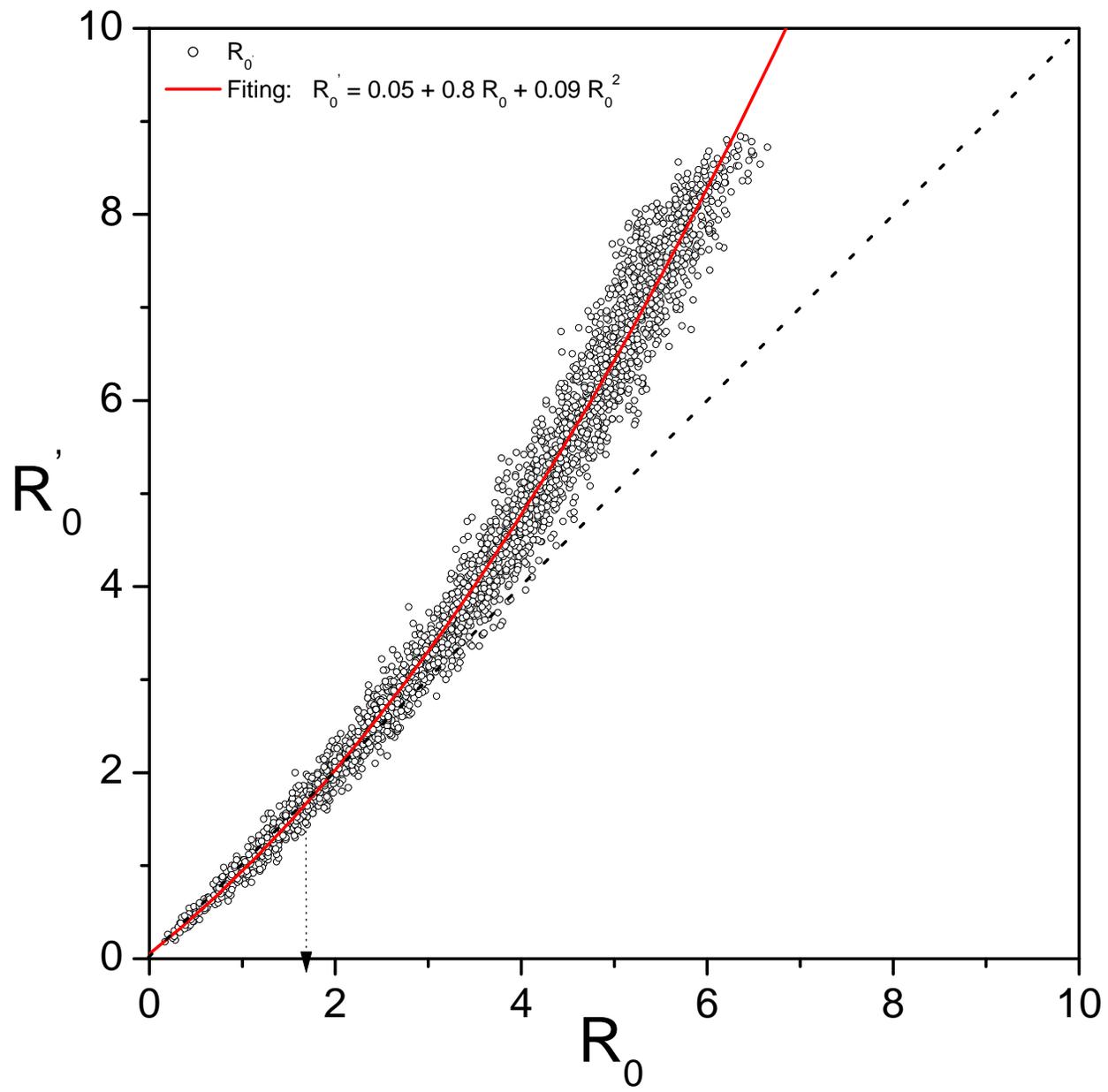
\end{document}